\documentclass[pra,twocolumn,notitlepage,showpacs]{revtex4-1}
\usepackage{graphicx}
\usepackage{amsmath}
\usepackage{amsfonts}
\usepackage{amssymb}
\usepackage{epsfig}
\usepackage[pdftex]{color}

\begin{document}

\title{Hybrid cavity mechanics with doped systems}

\author{Aur\'{e}lien Dantan$^1$, Bhagya Nair$^1$, Guido Pupillo$^2$, and Claudiu Genes$^3$}

\affiliation{$^1$Department of Physics and Astronomy, University of Aarhus, DK-8000 Aarhus C, Denmark\\
$^2$IPCMS (UMR 7504) and ISIS (UMR 7006), University of Strasbourg and CNRS (UMR 7006), 67000 Strasbourg, France\\
$^3$Institute for Theoretical Physics, University of Innsbruck, Technikerstrasse 25, A-6020 Innsbruck, Austria}

\date{\today}

\begin{abstract}
We investigate the dynamics of a mechanical resonator in which is embedded an ensemble of two-level systems interacting with an optical cavity field. We show that this hybrid approach to optomechanics allows for enhanced effective interactions between the mechanics and the cavity field, leading for instance to ground state cooling of the mechanics, even in regimes, like the unresolved sideband regime, in which standard radiation pressure cooling would be inefficient.
\end{abstract}

\pacs{42.50.Wk,42.50.Ct,85.85.+j,42.50.Pq}


\maketitle

\section{Introduction}
Functionalizing mechanical resonators in order to enhance their
response to electromagnetic fields has enabled numerous applications
such as force sensing in metrology or signal transduction and
storage in telecommunication sciences. With the recent advances in
the field of quantum optomechanics~\cite{Aspelmeyer2013},
\textit{hybrid} opto- or electromechanical architectures interfacing
atom or atomic-like systems with mechanical resonators represent an
interesting platform for investigating the coupling of
electromagnetic fields with mechanical motion at the quantum level.
The rich nature of interactions between electromagnetic radiation,
atomic systems and mechanics can be exploited for enhanced state
preparation, readout or transfer between disparate physical
systems~\cite{Treutlein2012,Wallquist2009}. Various hybrid
interfaces have been studied, in which single atoms or
molecules~\cite{Tian2004,Favero2008,Singh2008,Hammerer2009b,Puller2013,Restrepo2014},
cold atomic
ensembles~\cite{Meiser2006,Treutlein2007,Genes2008,Ian2008,Genes2009,Hammerer2009a,Bhattacherjee2009,Hammerer2010,Hunger2010,Paternostro2010,Camerer2011,Genes2011,Vogell2013},
quantum dots~\cite{WilsonRae2004,Lambert2008,Yeo2014}, NV
centers~\cite{Rabl2009,Arcizet2012,Kolkowitz2012,Zhang2013,Scala2013},
defects~\cite{Ramos2013}, artificial superconducting
atoms~\cite{OConnell2010,Pirkkalainen2013,Palomaki2013}, can
interact with movable mirrors, membranes, cantilevers, nanobeams,
etc.

A prototypical hybrid optomechanical system, as considered e.g.
in~\cite{Favero2008,Genes2008,Genes2009,Restrepo2014}, consists in a
single optical mode coupled, on the one hand, to a single mechanical
mode via radiation pressure and, on the other hand, to a single (or
ensemble of) two-level system (TLS). The interaction of a TLS with
the field of the optical resonator indirectly modifies the
optomechanical response of the mechanics, which may allow for
enhanced optomechanical cooling, coherent atom-photon-phonon
interactions or the generation of multipartite nonclassical states.

We propose here an alternative hybrid optomechanics approach in
which an ensemble of TLS which interacts with an optical cavity field is embedded directly into a macroscopic
mechanical resonator. We
show that the TLS effectively mediate interactions between
the cavity field and the mechanics, which may result, for instance,
in efficient cooling of the mechanics to the ground state, even in
the unresolved sideband regime where standard radiation pressure
cooling would be inefficient.

This approach may have several advantages: on the one hand, the TLS
can provide narrow resonances and, thereby, a sharper dispersive
optomechanical response for the mechanics. On the other hand, their
integration into a massive resonator allows for operation in a
highly-localized regime with respect to the optical field spatial
period, and consequently, an enhanced light-matter interaction.
This localization may allow e.g. for the generation or
detection of large quantum superposition states~\cite{Zhang2013,Scala2013}. Fundamentally, this strategy thus potentially paves the way towards the
realization of novel types of hybrid optomechanical interactions
which can be exploited to extend the degree of control of the
mechanics. Practically, it is naturally implementable for a wide
range of resonators, such as nanomembranes, microspheres or
cantilevers, and particularly relevant for resonators whose direct
coupling to light via radiation pressure is weak.

The paper is outlined as follows: Sec.~\ref{sec:model} presents the
model, the effective interaction Hamiltonian and the calculation of
the steady state covariance matrix following a standard linearized
treatment. In Sec.~\ref{sec:results} an analytical derivation of the
effective mechanical susceptibility and noise terms affecting the
mechanics is provided, followed by a discussion of various regimes
of interest and numerical results.

\section{Model}\label{sec:model}

\begin{figure}
\centering
\includegraphics[width=0.98\columnwidth]{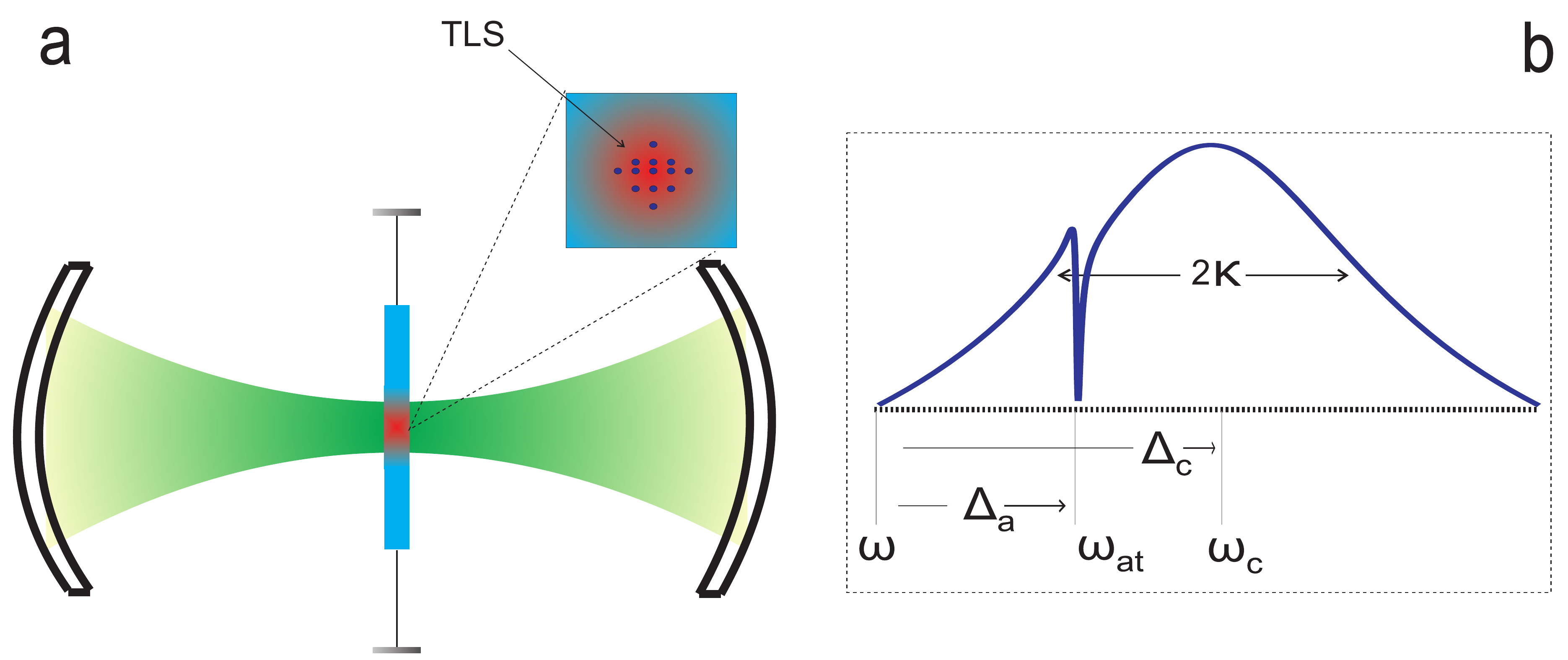}
\caption{\emph{Schematics }- (a) Hybrid optomechanical setup
consisting of a vibrating membrane inside an optical cavity
and doped with two-level quantum emitters (TLS) which interact with one cavity field mode. The doping pattern can be tailored to fit vibrational
patterns of the membrane and/or the transverse intensity profile of
the cavity field. (b) Cavity field susceptibility as a function of frequency, illustrating the bad cavity/good dopant regime in which the cavity linewidth $\kappa$ is much larger than the dopant linewidth $\gamma$. The relevant frequencies and detunings are shown (see text for details).} \label{fig:fig1}
\end{figure}

We consider the situation depicted in Fig.~\ref{fig:fig1}, in which a flexible membrane with thickness smaller than the relevant optical wavelengths is positioned inside an optical cavity~\cite{Thompson2008}. Owing to its intrinsic elastic properties and the boundary conditions imposed by the clamping geometry,
the membrane exhibits a set of normal modes, denoted by the index $s$,
with effective
frequencies $\omega _{\mathrm{s}}$, effective masses $m_{s}$ and displacement fields $u_{s}(\mathbf{r})$. We consider its motion along the (cavity) $x$-axis and denote the transverse
position in the ($y,z$)-plane by the vector $\mathbf{r}$. After
quantization, a general displacement operator in the direction of
interest $x$ can be expanded in terms of normal modes
\begin{equation}
    \hat{x}(\mathbf{r})=\sum_{\mathrm{s}} x_{\text{zpm}}^{\mathrm{s}} u_{\mathrm{s}}(\mathbf{r})(\hat{b}_{\mathrm{s}}+\hat{b}_{\mathrm{s}}^{\dagger }),
\end{equation}
where $\hat{b}_{\mathrm{s}}$ is the phonon annihilation operator
for mode $\mathrm{s}$, while its effective mass is given by
$m_{\mathrm{s}}=\rho \int d^{2}r\left\vert
u_{\mathrm{s}}(\mathbf{r})\right\vert ^{2}$ (assuming constant
surface density $\rho$). The zero-point amplitude is defined as
$x_{\text{zpm}}^{\mathrm{s}}=\sqrt{\hbar /\left(
m_{\mathrm{s}}\omega _{\mathrm{s}}\right) }$. For typical membranes vibrating at MHz frequencies with $\sim$ng effective mass it is of the order of
fm and one can reasonably assume localization deep into the
Lamb-Dicke limit even at room temperature.

We consider a poorly reflecting membrane whose dispersive optomechanical coupling with the cavity field is weak [this point is discussed further in Sec.~\ref{sec:discussion}]. We assume however that it is patterned with an ensemble $N$ two-level emitters whose density is low enough such the bare mechanical properties of the membranes are unchanged, but which interacts with the cavity field with dynamics described by the standard Tavis-Cummings
interaction. The $N$ two-level systems, with
internal transition described by Pauli operators $\hat{\sigma}_{+}^{(j)}$
and $\hat{\sigma}_{-}^{(j)}$, are located at a set of positions
$\{\mathbf{r}_{j},x_{j}\}$, with $j$ from $1$
to $N$, according to a surface distribution function $h(\mathbf{r})\delta (%
\mathbf{r}-\mathbf{r}_{j})$; in principle, they can interact with a multitude of
cavity field modes with annihilation operators $a_{n}$, frequency
$\omega _c^{(n)}$ and spatial structure assumed of the form
$f_{n}(\mathbf{r})t(x)$. The
Tavis-Cummings interaction can then be written as
\begin{equation}
H_{\mathrm{int}}=\sum_{n,j}\int d\mathbf{r}g_{0}\left[ f_{n}(\mathbf{r})h(\mathbf{r}%
)t(\hat{x}(\mathbf{r}))\hat{\sigma}_{+}^{(j)}\hat{a}_{n}+h.c.\right] \delta (\mathbf{r%
}-\mathbf{r}_{j}),
\label{hamgen}
\end{equation}%
where $g_{0}=d\sqrt{\hbar \omega _c^{(n)}/2\epsilon _{0}V}$ is the vacuum Rabi frequency, $d$ is the
transition dipole moment, $\epsilon _{0}$ the permittivity of the
vacuum and $V$ the quantization volume.

Let us assume a fixed equilibrium position $x_{0}(\mathbf{r})$ of the membrane and look at the coupling of its vibrations along the $x$-axis to
the light field via the TLS. One can perform the following expansion of any function of position: $t(\hat{x}(\mathbf{r}%
))\simeq t(x_{0}(\mathbf{r}))+t^{\prime }(x_{0}(\mathbf{r}))\hat{x}(%
\mathbf{r})+...$. The integral $\hat{S}_{+}^{(0,n)}=\sum_{j}\int d%
\mathbf{r}f_{n}(\mathbf{r})h(\mathbf{r})t(x_{0}(\mathbf{r}))\delta (\mathbf{r}-%
\mathbf{r}_{j})\hat{\sigma}_{+}^{(j)}$ and its hermitian conjugate are identified as the
\textit{static} contribution to the interaction due to the presence of collective spin operators $\hat{S}_{+}^{(0,n)}$ and $\hat{S}_{-}^{(0,n)}$, while the \textit{dynamical} contribution involves the collective operators $\hat{S}_{+}^{(1,s,n)}=\sum_{j}\int
d\mathbf{r}f_{n}(\mathbf{r})h(\mathbf{r})t^{\prime
}(x_{0}(\mathbf{r}))u_{s}(\mathbf{r})\hat{\sigma}_{+}^{(j)}$ and its hermitian conjugate $\hat{S}_{-}^{(1,s,n)}$. The first part of the Hamiltonian simply describes the static atom-light interaction situation, $\sum_{n}g_{0}(\hat{S}_{+}^{(0,n)}\hat{a}_{n}+h.c.)$, for a collection of $N$ atoms
described by a collective spin operator $\hat{S}_{+}^{(0)}$. The second term, linear in the position operator, takes the form $\sum_{n,\mathrm{s}}g_{0}\beta_{\mathrm{s}}
\left(\hat{S}_{+}^{(1,\mathrm{s},n)}\hat{a}_{n}+h.c.\right)(\hat{b}_{\mathrm{s}}+\hat{b}_{\mathrm{s}}^{\dagger })$, where, for each
vibrational mode, the overlap between the doping pattern, the field
mode spatial structure and the vibrational mode transverse
oscillation profile defines the addressed collective spin operators.

\subsection{Simplified model}

Such a hybrid optomechanical system provides in principle a multitude of
operational points defined by diverse mechanical and field mode
patterns. We will now restrict our investigations to the situation where the membrane doping pattern matches a chosen
vibrational mode. The cavity mode at frequency $\omega_c$ is
driven optically via the side mirror with a laser of amplitude
$\eta$ and frequency $\omega$. Moreover, we assume a weak driving
of the TLS ensemble, which allows
us to perform the usual linearization via the Holstein-Primakoff transformation $\hat{S}_{z}=-N/2+\hat{c}^{\dagger }\hat{c}$ and $\hat{S}_{-}=\sqrt{N}\hat{c}$. In this
simplified picture the interaction Hamiltonian reads
($\hbar=1$)
\begin{equation}\label{eq:Hint}
H_{\mathrm{int}}=g(\hat{q},q_0)(\hat{a}^{\dagger}\hat{c}+\hat{a}\hat{c}^{\dagger }),
\end{equation}%
where the position dependent coupling $g(\hat{q},q_0)$ includes the
overlap integrals in Eq.~(\ref{hamgen}) and is written in terms of
the position quadrature $\hat{q}=(\hat{b}+\hat{b}^{\dagger})/\sqrt{2}$. The collective coupling scales as $g_{0}\sqrt{N}$, where $N$ represents the effective number of emitters involved in the interaction, given by the
overlap integrals previously discussed. The free evolution
(in a frame rotating at $\omega$) is governed by
\begin{equation}\label{eq:H0}
H_0=\frac{1}{2}\omega_m(\hat{p}^2+\hat{q}^2)+\Delta_c\hat{a}^{\dagger}\hat{a}+\Delta_a\hat{c}^{\dagger}\hat{c},
\end{equation}
where $\hat{p}=i(\hat{b}^{\dagger}- \hat{b})/\sqrt{2}$ and
$\Delta_c=\omega_c-\omega$, $\Delta_a=\omega_{at}-\omega$ are the
cavity and atomic detunings and $\omega_m$ the mechanical frequency (Fig.~\ref{fig:fig1}) and the conventions
$[\hat{a},\hat{a}^{\dagger}]=[\hat{c},\hat{c}^{\dagger}]=1$ and
$[\hat{q},\hat{p}]=i$ are used. The driving Hamiltonian is
\begin{equation}\label{eq:HL}
H_{\mathrm{L}}=i \eta (\hat{a}^{\dagger }-\hat{a})
\end{equation}

For a complete description of the dynamics the dissipation channels
have to be accounted for. These consist of i) losses owing to the
mechanical coupling to thermal environment at a
rate $\gamma _{m}$. ii) cavity losses of photons at a rate $\kappa $
and iii) TLS decay at a
rate $\gamma$. The corresponding Liouvillians are $\mathcal{L}_{\hat{O}}[\rho ]=\Gamma _{\hat{O}}D_{\hat{O}}[\rho
]=\Gamma_{\hat{O}} \lbrack \hat{O}\rho \hat{O}^{\dagger }-\hat{O}^{\dagger }\hat{O}\rho -\rho
\hat{O}^{\dagger }\hat{O}]$ with $\hat{O}$ and $\Gamma _{\hat{O}}$ standing for the collapse
operators $\hat{a},\hat{b},\hat{c}$ and their associated loss rates $\kappa ,\gamma _{m}$
and $\gamma$, respectively.

We furthermore assume a linear optomechanical coupling, i.e.
\begin{equation}
g(\hat{q},q_0)=g^{(0)}+g^{(1)}\hat{q}
\end{equation}
where the static and linear couplings are $g^{(0)}=g_0\sqrt{N}\sin(kq_0)$,
$g^{(1)}=g_0\sqrt{N}\beta\eta_{\textrm{LD}}\cos(kq_0)$ with $\eta_{\textrm{LD}}=kx_{\text{zpm}}$ the Lamb-Dicke parameter and $\beta$ a geometrical factor coming from the overlap integrals previously defined. We choose a position $q_0$ of the
membrane in the cavity such that \textit{both} $g^{(0)}$ and $g^{(1)}$ are
non-zero.

\subsection{Equations of motion}

Adding dissipation to the full Hamiltonian evolution given by Eqs.~(\ref{eq:Hint},\ref{eq:H0},\ref{eq:HL}) yields the
following Heisenberg-Langevin equations of motion
\begin{eqnarray}
\dot{\hat{a}}&=&-(\kappa+i\Delta_c)\hat{a}-i(g^{(0)}+g^{(1)}\hat{q})\hat{c}+\eta+\hat{a}_{in}\\
\dot{\hat{c}}&=&-(\gamma+i\Delta_a)\hat{c}-i(g^{(0)}+g^{(1)}\hat{q})\hat{a}+\hat{c}_{in}\\
\dot{\hat{p}}&=&-\gamma_m \hat{p}-\omega_m \hat{q}-g^{(1)}(\hat{a}^{\dagger}\hat{c}+\hat{c}^{\dagger}\hat{a})+\hat{\xi}\\
\dot{\hat{q}}&=&\omega_m\hat{p}
\end{eqnarray}
where the field, atomic and mechanical input noise terms, $\hat{a}_{in}$, $\hat{c}_{in}$ and $\hat{\xi}$, are zero mean-valued and have correlation functions $\langle\hat{a}_{in}(t)\hat{a}^{\dagger}_{in}(t') \rangle=2\kappa\delta(t-t')$, $\langle\hat{c}_{in}(t)\hat{c}^{\dagger}_{in}(t') \rangle=2\gamma\delta(t-t')$ and $\langle\hat{\xi}(t)\hat{\xi}(t') \rangle=\gamma_m(1+2n_m)\delta(t-t')$, respectively. For the thermal noise term, the standard Ohmic bath approximation has been made and $n_m$ represents the initial thermal occupation number of the mechanical mode considered.

\subsection{Steady state and linearized equations}
Denoting by $\bar{O}=\langle O\rangle$ the steady state value of $\hat{O}$, and using $\bar{c}_{in}=0$, $\bar{a}_{in}=0$ and $\bar{\xi}=0$, one obtains the steady state mean values by solving the following set of equations
\begin{eqnarray}
0&=&-(\kappa+i\Delta_c)\bar{a}-i(g^{(0)}+g^{(1)}\bar{q})\bar{c}+\eta\\
0&=&-(\gamma+i\Delta_a)\bar{c}-i(g^{(0)}+g^{(1)}\bar{q})\bar{a}\\
0&=&-\omega_m \bar{q}-g^{(1)}(\bar{a}^*\bar{c}+\bar{c}^*\bar{a})
\end{eqnarray}
which gives
\begin{equation}
\bar{c}=-\frac{i(g^{(0)}+g^{(1)}\bar{q})}{\gamma+i\Delta_a}\bar{a}\equiv-\frac{ig}{\gamma+i\Delta_a}\bar{a}
\end{equation}
where, in order to make the connection with the standard cavity QED settings, we included the mean position shift in an effective atom-light coupling $g=g^{(0)}+g^{(1)}\bar{q}$.
Typically, $g^{(1)}$ is small enough so that $g\simeq g^{(0)}$ actually represents the standard collective coupling rate for a superemitter localized at position $q_0$. For not too high intracavity photon numbers, one can also reasonably assume that the optical spring-induced modification of the mechanical frequency is such that $\omega_m>2g^{(1)}|\bar{a}|^2\Delta_a/(\gamma^2+\Delta_a^2)$. The mean position shift is then given by
\begin{equation}
\bar{q}=\frac{2g^{(1)}|\bar{a}|^2\frac{g^{(0)}\Delta_a}{\gamma^2+\Delta_a^2}}{\omega_m-\frac{2\Delta_aG^2}{\gamma^2+\Delta_a^2}},
\end{equation}
where we defined $G=g^{(1)}\bar{a}$ in analogy with the enhanced optomechanical coupling in the standard dispersive optomechanics in the linearized regime (see Sec.~\ref{sec:standard}). $\bar{a}$ is solution of
\begin{equation}
\left[\kappa+i\Delta_c+\frac{g^2}{\gamma+i\Delta_a}\right]\bar{a}=\eta.
\end{equation}
Without loss of generality one can assume $\bar{a}$ real and positive. Let us note that, at high optomechanical coupling strengths, the previous equation may give rise to multistable solutions for the intracavity field photon number, since $g$ depends on $\bar{a}$ through $\bar{q}$. We assume in the following that we operate outside of this instability regime.

Assuming these static stability conditions met, we proceed with the usual linearization around steady state by decomposing each observable as the sum of its steady state mean values and its fluctuations $\hat{o}=\bar{o}+o$. Neglecting second order terms for the fluctuations, we obtain
\begin{eqnarray}
\dot{a}&=&-(\kappa+i\Delta_c)a+G\frac{g}{\gamma+i\Delta_a}q+a_{in}\label{eq:dota}\\
\dot{c}&=&-(\gamma+i\Delta_a)c-iga-iGq+c_{in}\\
\dot{p}&=&-\gamma_mp-\omega_m q-G\left[c+\frac{ig}{\gamma-i\Delta_a}a+\textrm{h.c.}\right]+\xi\\
\dot{q}&=&\omega_mp\label{eq:dotq}
\end{eqnarray}
The dynamic stability of the system can thus be determined by examining the eigenvalues of the evolution matrix
\begin{equation}
[A]=\left(\begin{array}{cccccc}
-\gamma & \Delta_a & 0 & g & 0 & 0\\
-\Delta_a & -\gamma & -g & 0 & -\sqrt{2}g & 0\\
0 & g & -\kappa & \Delta_c & -\frac{gG\gamma\sqrt{2}}{\gamma^2+\Delta_a^2} & 0\\
-g & 0 & -\Delta_c & -\kappa & \frac{gG\Delta_a\sqrt{2}}{\gamma^2+\Delta_a^2} & 0\\
0 & 0 & 0 & 0 & 0 & \omega_m\\
-G\sqrt{2} & 0 & -\frac{igG}{\gamma-i\Delta_a} & \frac{igG}{\gamma+i\Delta_a} & -\omega_m & -\gamma_m
\end{array}\right)
\end{equation}
expressed in the basis of the quadratures $(X,Y,x,y,q,p)$, where $X=(c+c^{\dagger})/\sqrt{2}$, $Y=i(c^{\dagger}-c)/\sqrt{2}$, $x=(a+a^{\dagger})/\sqrt{2}$, $y=i(a^{\dagger}-a)/\sqrt{2}$.
When the real part of each eigenvalue is strictly negative the system is stable and the steady state covariance matrix of the system, $[V]$, can be calculated by solving the Lyapunov equation
\begin{equation}
[A] [V]+[V][A]^{\dagger}=-[D]
\end{equation}
where $[D]=\textrm{diag}[\gamma,\gamma,\kappa,\kappa,0,\gamma_m(1+2n_m)]$ is the diffusion matrix.

\subsection{Effective mechanical susceptibility, noise spectrum and final occupation number for the mechanics}
In order to get some more insight into the dynamics of the system we analytically derive the effective mechanical susceptibility and the effective noise terms for the mechanics. To do so, we Fourier transform eqs.~(\ref{eq:dota}-\ref{eq:dotq}) and, after some algebra, obtain the expression of the Fourier transform of the position fluctuations $q(\omega)$
\begin{align}
\label{eq:qomega} &\chi_m^{\textrm{eff}}(\omega)^{-1} q(\omega)\\
\nonumber &=\Lambda(\omega)c_{in}+\Lambda^*(-\omega)c_{in}^{\dagger}+\Upsilon(\omega)a_{in}+\Upsilon^*(-\omega)a_{in}^{\dagger}+\xi
\end{align}
where $\chi_m^{\textrm{eff}}(\omega)^{-1}$ is the effective mechanical susceptibility
\begin{equation}\chi_m^{\textrm{eff}}(\omega)^{-1}=\chi_m(\omega)^{-1}+\Theta(\omega)+\Xi(\omega)\label{eq:chieff}\end{equation}
with
\begin{align}
\chi_m(\omega)^{-1}&=(\omega_m^2-\omega^2-i\gamma_m\omega)/\omega_m\\
\Theta(\omega)&=-\frac{2G^2\Delta_a}{(\gamma-i\omega)^2+\Delta_a^2}\label{eq:Theta}\\
\Xi(\omega)&=-G^2[\chi_c(\omega)A(\omega)+\chi_c^*(-\omega)A^*(-\omega)]\label{eq:Xi}\\
\chi_c(\omega)^{-1}&=\kappa+i\Delta_c-i\omega+\frac{g^2}{\gamma+i\Delta_a-i\omega}\\
A(\omega)&=\frac{ig^2}{\gamma^2+\Delta_a^2}\frac{(2\gamma+2i\Delta_a-i\omega)(2i\Delta_a-i\omega)}{(\gamma+i\Delta_a-i\omega)^2}
\end{align}
and
\begin{align}
\Lambda(\omega)&=\frac{G}{\gamma+i\Delta_a-i\omega}\left[1+\frac{g^2}{\gamma-i\Delta_a}\chi_c(\omega)B(\omega)\right]\\
\Upsilon(\omega)&=G\frac{ig}{\gamma-i\Delta_a}\chi_c(\omega)B(\omega)\\
B(\omega)&=\frac{2i\Delta_a-i\omega}{\gamma+i\Delta_a-i\omega}
\end{align}
Eq.~(\ref{eq:qomega}) shows that the mechanical oscillator fluctuations are given by the product of the effective susceptibility and the sum of fluctuations arising from three uncorrelated noise terms coming from the atoms, the incoming field and the coupling with the thermal reservoir, respectively. One can compute the steady state noise spectrum of the position observable by
\begin{equation}
S_q(\omega)=|\chi_m^{\textrm{eff}}(\omega)|^2\left[2\gamma|\Lambda(\omega)|^2+2\kappa|\Upsilon(\omega)|^2+\gamma_m(1+2n_m)\right]
\end{equation}
As a figure of merit for cooling we will consider the final occupation number in the mechanics obtained by integration of the noise spectrum of the position observable
\begin{equation}
n_f=\Delta q^2-\frac{1}{2}=\int\;\frac{d\omega}{2\pi}S_q(\omega)-\frac{1}{2}
\end{equation}
Note that, strictly speaking, this occupation number should be defined as $(\Delta q^2+\Delta p^2-1)/2$~\cite{Genes2008b}, but, for the situations we will consider, the difference is negligible.

\subsection{Standard dispersive radiation pressure optomechanics}\label{sec:standard}
Before exploring the dynamics of the doped system in various parameter regimes it is interesting to briefly recall the results for the standard radiation pressure optomechanics in the dispersive regime. Starting from the Hamiltonian $H_{OM}=G_0\hat{a}^{\dagger}\hat{a}\hat{q}$ and assuming a weak single-photon optomechanical coupling $G_0$, the same linearization approach would yield an effective mechanical susceptibility~\cite{WilsonRae2007,Marquardt2007,Dantan2008,Genes2008b}
\begin{equation}\label{eq:standardchi}
\chi_m^{\textrm{OM}}(\omega)^{-1}=\chi_m(\omega)^{-1}-\frac{2G_{\textrm{OM}}^2\Delta_c}{(\kappa-i\omega)^2+\Delta_c^2}
\end{equation}
with $G_{\textrm{OM}}=G_0\bar{a}$. For a high mechanical quality factor, the mechanics noise spectrum is still approximately that of a harmonic oscillator, but with an effective mechanical damping modified by the radiation pressure
\begin{equation}\label{eq:gamma_mOM}
\gamma_m\rightarrow\gamma_m+\Im\left[\frac{2G_{\textrm{OM}}^2\Delta_c}{(\kappa-i\omega_m)^2+\Delta_c^2}\right]
\end{equation}
In the good cavity limit, $\kappa\ll\omega_m$, driving the mechanics on the red sideband ($\Delta_c=\omega_m$) gives cooling of the mechanics with a rate $\Gamma=G_{\textrm{OM}}^2/\kappa$, while being resonant with the blue sideband can give rise to self-oscillations when $G_{\textrm{OM}}^2\gtrsim 2\kappa\gamma_m$. In the bad cavity limit, $\kappa\gg\omega_m$, cooling is optimum for $\Delta_c\sim \kappa/\sqrt{3}$, while self-oscillations also occur as soon as $G_{\textrm{OM}}^2\gtrsim 2\kappa\gamma_m$ for $\Delta_c\sim -\kappa$. Ground state cooling is possible in the good cavity limit where, for $\Delta_c=\omega_m$ and neglecting second order terms in $(\kappa/\omega_m)^2$, the final occupation number is given by
\begin{equation}\label{eq:nfOM}
n_f=\frac{\gamma_m}{\gamma_m+\Gamma}n_m
\end{equation}

\subsection{Doped versus radiation pressure optomechanics}\label{sec:discussion}
In principle, both the dopant-mediated coupling and the radiation pressure forces can affect the mechanics of the semi-transparent membrane. It is thus interesting to compare their magnitude in the linearized regime considered here. For a membrane with amplitude reflectivity coefficient $r$, the single-photon optomechanical coupling in the standard dispersive optomechanics scenario discussed in the previous section is given by $G_0=r(\omega/L)x_{\textrm{zpm}}$, where $L$ is the cavity length. Including the term arising from the Hamiltonian $H_{OM}$ in the dynamical equation for the intracavity field fluctuations (\ref{eq:dota}) would modify it to
\begin{equation}\dot{a}=-(\kappa+i\Delta_c)a+\left[G\frac{g}{\gamma+i\Delta_a}-iG_{\textrm{OM}}\right]q+a_{in}\label{eq:dota2}\end{equation}
where $G=g^{(1)}\bar{a}$ and $G_{\textrm{OM}}=G_0\bar{a}$. Anticipating on the results in the next section, we set $\Delta_a=0$ and compare the ratio of the modulus of the two terms in the square brackets: $(Gg/\gamma)/G_{\textrm{OM}}$, which essentially gives the ratio of the magnitude of the two optomechanical forces. Recalling that $g^{(1)}\sim g\eta_{\textrm{LD}}\beta= g(\omega/c)x_{\textrm{zpm}}\beta$ and assuming $\beta\sim 1$, this ratio becomes
\begin{equation}
\frac{gG/\gamma}{G_{\textrm{OM}}}\sim\frac{g^2L}{rc\gamma}=\frac{\alpha}{r}
\end{equation}
where $\alpha=g^2L/(c\gamma)=\frac{3}{4\pi}\frac{\lambda^2}{S}N$ is the single-pass optical depth of the dopant ensemble ($S$ being the beam cross section). This simple order of magnitude estimate shows that the doped optomechanical force can dominate over the dispersive radiation pressure force for poorly reflecting membranes and optically dense dopant.

In the following we will compare the effect of these forces acting separately on the mechanics, the extension to the situation where both forces simultaneously play a role being straightforward.

\section{Results}\label{sec:results}

\subsection{Effective resolved sideband cooling with good dopant}
Of particular interest is a dopant having a narrow resonance as compared to the cavity linewidth and the mechanical frequency, as it can allow for effective resolved sideband cooling of the mechanics.

\begin{figure}
\centering
\includegraphics[width=0.5\textwidth]{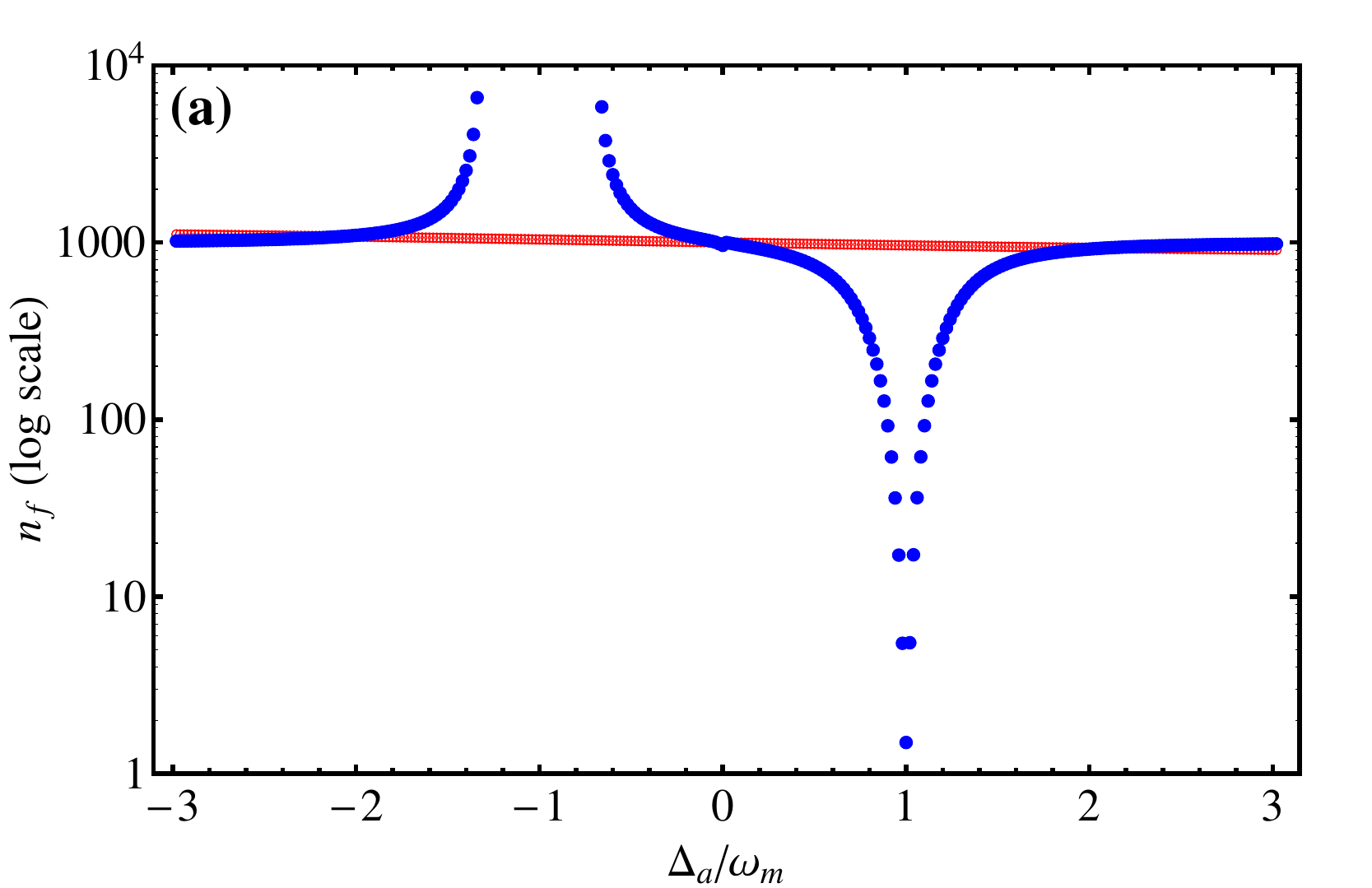}\\
\includegraphics[width=0.5\textwidth]{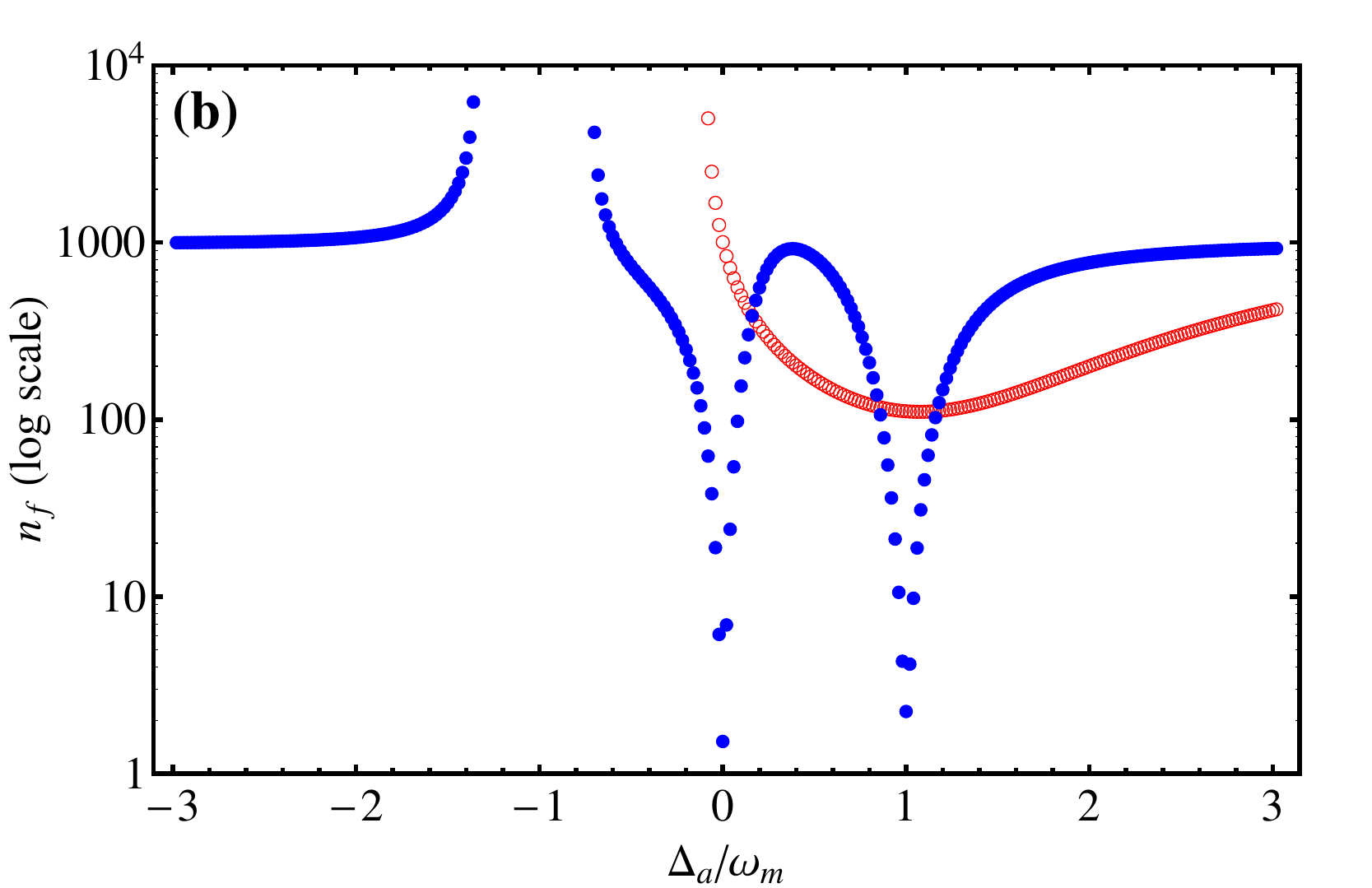}
\caption{Final occupation number $n_f$ as a function of atomic detuning $\Delta_a$ (blue dots): in (a) the bad ($\kappa/\omega_m=10$, $g/\omega_m=10^{-2}$) and (b) intermediate ($\kappa/\omega_m=1$, $g/\omega_m=0.1$) cavity regimes. Parameters: $\gamma/\omega_m=10^{-2}$, $\Delta_c/\omega_m=1$, $\gamma_m/\omega_m=10^{-5}$, $G/\omega_m=10^{-2}$, and $n_m=10^3$. The corresponding standard radiation pressure OM result (in which case the $x$-axis represents $\Delta_c/\omega_m$) is shown by the red circles for comparison. Only points corresponding to stable working points are shown.}
\label{fig:fig2}
\end{figure}

For a bad cavity, i.e. for a cavity field decay rate $\kappa$ much larger than the other relevant frequencies and rates, the field susceptibility $\chi_c(\omega) $ is small, which means that one can neglect $\Xi(\omega)$ in Eq.~(\ref{eq:chieff}), and the effective mechanical susceptibility is dominated by the atomic response $\Theta(\omega)$. The expression of $\Theta(\omega)$ [Eq.~(\ref{eq:Theta})] is formally identical to the field response in the standard theory for dispersive radiation pressure optomechanics, with the atoms replacing the field [i.e. replacing $\Delta_c$ by $\Delta_a$ and $\kappa$ by $\gamma$ in Eq.~(\ref{eq:standardchi})]. Indeed, for $\gamma,g\ll\omega_m\ll\kappa$ and neglecting $\Xi(\omega)$, driving the mechanics on the red sideband $\Delta_a=\omega_m$ gives rise to optomechanical cooling with a rate $\Gamma=G^2/\gamma$ and a cooling limit given by Eq.~(\ref{eq:nfOM}) in the effectively resolved sideband regime induced by the atoms. This is illustrated in Fig.~\ref{fig:fig2}a, which shows the final occupation number as a function of the atomic detuning for a 'good' dopant ($\gamma/\omega_m=0.01$, blue dots) in a 'bad' cavity ($\kappa/\omega_m=10$). For comparison, the red curve shows the result in the corresponding standard radiation pressure scenario (for the same cavity and comparable optomechanical coupling $G_{\textrm{OM}}=G$), which displays very inefficient cooling, as $\Gamma\sim\gamma_m/10$ in this case.

Let us now consider a cavity in the intermediate regime $\omega_m\sim\kappa\gg g$, for which one can no longer neglect the contribution from the field susceptibility in $\Xi(\omega)$. If one assumes that the cavity is also detuned to the red sideband ($\Delta_c=\omega_m$), one can neglect blue sideband contributions in the effective susceptibility as well as in the effective noise terms. Introducing the cooperativity parameter $C=g^2/\kappa\gamma$, one has $\chi_c(\omega_m)\sim 1/(\kappa+g^2/\gamma)=1/\kappa(1+C)$ and the optomechanical damping becomes
\begin{equation}
\label{eq:Gamma_detuned}
\Gamma=\frac{G^2}{\gamma(1+C)}. \end{equation}
The effective atomic and field noise terms reduce to $[G/(1+C)]c_{in}$ and $[-igG/\kappa\gamma(1+C)]a_{in}$, respectively. It follows that the final occupation number is formally given by Eq.~(\ref{eq:nfOM}) with $\Gamma$ defined by Eq.~(\ref{eq:Gamma_detuned}). The increase in the effective cavity linewidth due to the coupling with the dopant thus effectively reduces the cooling rate.

The blue circles in Fig.~\ref{fig:fig2}b show the final occupation number as a function of the atomic detuning for a 'good' dopant in this intermediate cavity regime ($\omega_m=\kappa$). As expected, the dip occurring around $\Delta_a=\omega_m$ corresponds to an effective resolved sideband cooling of the mechanics, which is slightly less efficient than in Fig.~\ref{fig:fig2}a, although still noticeably better than the corresponding radiation pressure scenario (red dots). However, one also observes a second cooling 'dip', occurring for $\Delta_a=0$, which we discuss in the next section.

\subsection{Enhanced optomechanical interactions with resonant dopant}
As observed in the previous section another interesting regime for enhancing the optomechanical interaction is to have the driving laser frequency resonant ($\Delta_a=0$) with a doping medium which is in the resolved sideband regime ($\gamma\ll\omega_m$). In this case, the purely atomic contribution $\Theta(\omega)$ vanishes and only the fluctuations of the cavity field (dressed by the dopant) contribute to the effective mechanical susceptibility. With $A(\omega_m)\simeq i(g/\gamma)^2$, the effective mechanical damping is given by
\begin{equation}\label{eq:gamma_mresonant}
\gamma_m+\Im\left[\frac{2G^2\left(\frac{g}{\gamma}\right)^2\Delta_c}{\left(\kappa-i\omega_m+\frac{g^2}{\gamma-i\omega_m}\right)^2+\Delta_c^2}\right]
\end{equation}
This result is again reminiscent of the standard dispersive optomechanics result [Eq.~(\ref{eq:gamma_mOM})]. However, a first noticeable difference is that the optomechanical coupling rate $G$ is now multiplied by the ratio $g/\gamma$, which can be substantially larger than unity for a strongly coupled dopant. The other noteworthy difference is that the cavity susceptibility is dressed by the dopant, as evidenced by the term $g^2/(\gamma-i\omega_m)$ in the denominator of Eq.~(\ref{eq:gamma_mresonant}). Still assuming that $\gamma\ll\omega_m$, this implies that the sign and amplitude of the second term in Eq.~(\ref{eq:gamma_mresonant}) essentially depend on the quantity $\omega_m-g^2/\omega_m$. Two regimes can then be distinguished: (i) $g\ll\omega_m$: a 'weak' coupling regime with the dopant, for which optimal cooling is obtained with a cavity field tuned to the red sideband $\Delta_c\sim\omega_m$, and (ii) $g\gg\omega_m$: a 'strong' coupling regime with the dopant, for which cooling can be achieved with a blue detuned cavity field.

More precisely, if one imposes
\begin{equation}\Delta_c=\Delta_0\equiv\omega_m-\frac{g^2}{\omega_m},\end{equation}
one has that
\begin{equation}\Xi(\omega_m)\simeq -G^2\left(\frac{g}{\gamma}\right)^2\frac{2\Delta_0}{\kappa^2-2i\kappa\Delta_0}\end{equation}

In the good cavity limit ($\kappa\ll|\Delta_0|$), the effective cooling rate is given by \begin{equation}\label{eq:Gammaenhanced}\Gamma=\frac{G^2}{\kappa}\left(\frac{g}{\gamma}\right)^2 \end{equation}
As aforementioned, it is enhanced with respect to the standard rate by the factor $(g/\gamma)^2$. Moreover, one has $B(\omega)\simeq 1$, $\Lambda(\omega_m)\simeq \frac{iG(1+g^2/\kappa\gamma)}{\omega_m}$ and $\Upsilon(\omega_m)\simeq \frac{igG}{\gamma\kappa}$, which yields a final occupation number given by Eq.~(\ref{eq:nfOM}), plus an extra atomic noise contribution equal to $\frac{(1+C)^2}{C}\left(\frac{\gamma}{\omega_m}\right)^2$. This shows that ground state cooling is in principle possible, but, since $C$ increases with $g$, a too strong coupling with the dopant may increase the amount of added atomic noise. There is therefore a tradeoff between enhanced cooling rate and added extra atomic noise.

\begin{figure}
\centering
\includegraphics[width=0.5\textwidth]{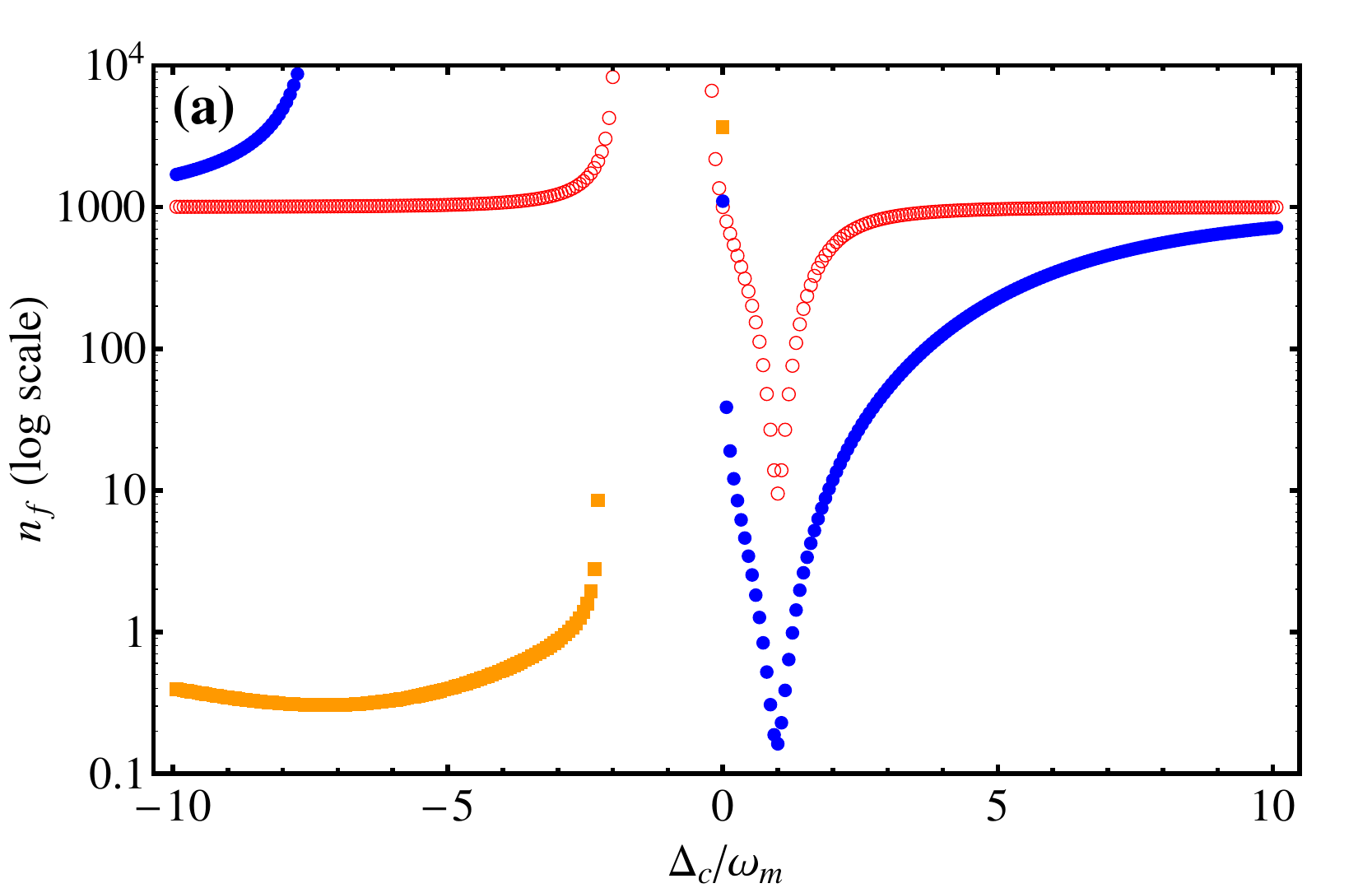}\\
\includegraphics[width=0.5\textwidth]{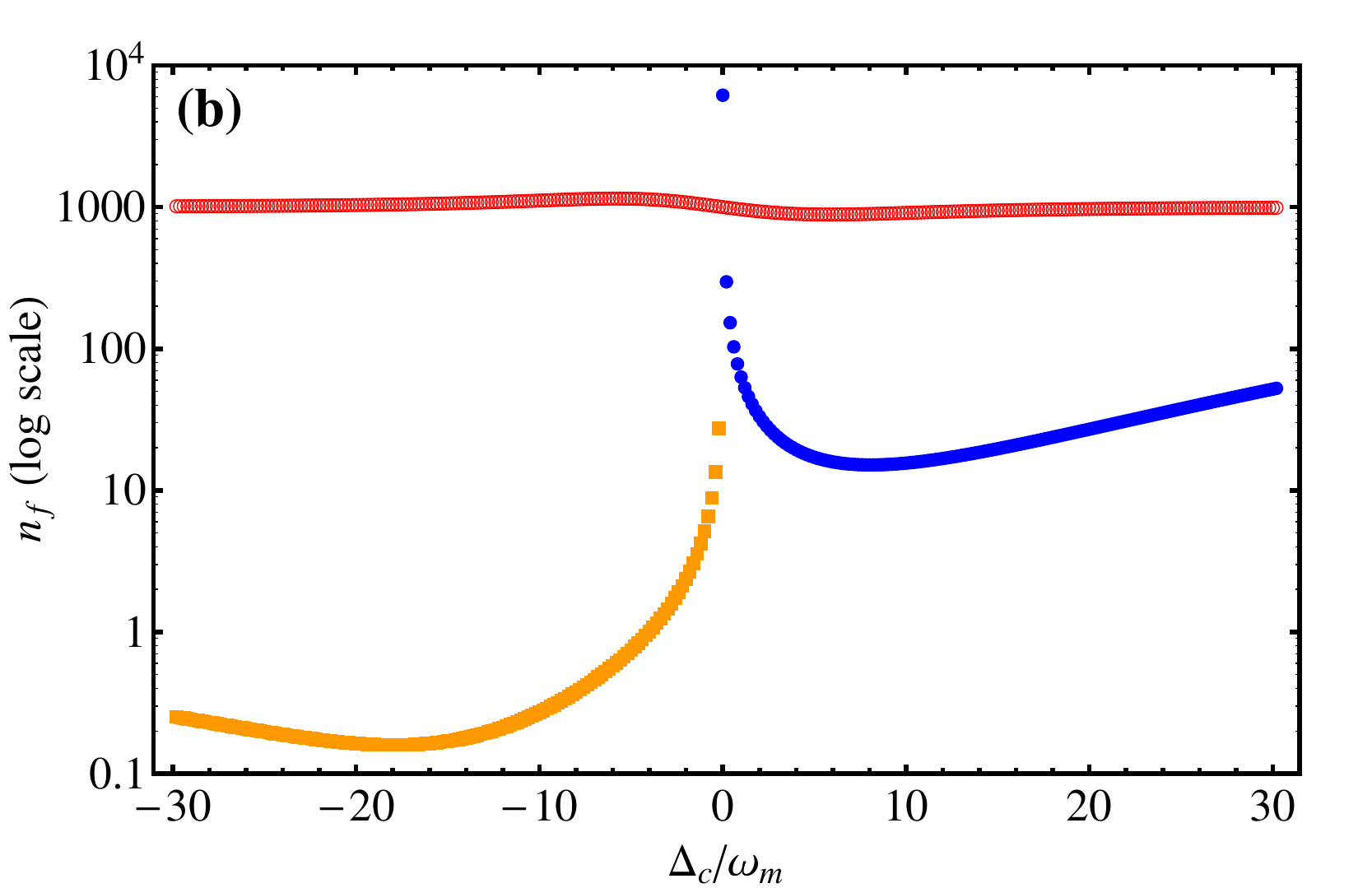}
\caption{Final occupation number $n_f$ as a function of cavity detuning $\Delta_c$ for a resonant dopant ($\Delta_a=0$).\\ (a) \textit{Good cavity limit}  ($\kappa/\omega_m=0.1$) -- Red circles: Standard OM. Blue dots: doped OM with a 'weakly' coupled dopant $g/\omega_m=0.1$. Yellow squares: doped OM with a 'strongly' coupled dopant $g/\omega_m=3$. (b) \textit{Bad cavity limit} ($\kappa/\omega_m=10$) -- Red circles: Standard OM. Blue dots: doped OM with a 'weakly' coupled dopant $g/\omega_m=0.72$. Yellow squares: doped OM with a 'strongly' coupled dopant $g/\omega_m=4$. Other parameters: $\gamma_m/\omega_m=10^{-5}$,  $G/\omega_m=10^{-2}$, $\gamma/\omega_m=10^{-2}$ and $n_m=10^3$. Only points corresponding to stable working points are shown.}
\label{fig:fig3}
\end{figure}

This result is illustrated in Fig.~\ref{fig:fig3}a, in which the final occupation number as a function of the cavity detuning is shown in both regimes, $g/\omega_m=0.1$ (blue dots) and $g/\omega_m=3$ (yellow squares). The bare optomechanical coupling rate and parameters were chosen so that standard OM cooling (red circles) with an equivalent $G$ does not allow for reaching the motional ground state. Clearly, an improvement of about two orders of magnitude is possible for a weakly coupled doped system around $\Delta_c\sim\omega_m$. As a result of the enhanced optomechanical coupling, optomechanical instabilities for blue cavity detunings also occur comparatively sooner than in the standard OM situation. In the strong coupling regime (yellow squares), as expected from the previous discussion, instabilities are observed for red detunings, while efficient cooling to the ground state is achieved for a wide range of blue detunings around $\Delta_c\sim\Delta_0\simeq-8\omega_m$.

In the bad (dressed) cavity limit ($\kappa\gg|\Delta_0|$) and for a cavity detuned by $\Delta_c\sim\kappa$, the cooling rate is given by \begin{equation}\Gamma=\frac{G^2}{\kappa}\left(\frac{g}{\gamma}\right)^2\frac{\omega_m}{\kappa}. \end{equation} Similarly to the standard OM result, the cooling rate is decreased by a factor $\sim\omega_m/\kappa$ as compared to the good cavity limit, and the lowest achievable final occupation number of Eq.~(\ref{eq:nfOM}) has to be divided by the same factor. However, let us note that, in the strongly coupled dopant regime $g>\omega_m$, $\kappa$ should be compared to $g^2/\omega_m$ in the dressed susceptibility. This means that one can operate in the unresolved sideband regime for the bare system ($\omega_m<\kappa$), but effectively be in the resolved sideband regime for a strongly coupled doped system satisfying $g^2>\omega_m\kappa$, and thereby still achieve ground state cooling.

This is illustrated in Fig.~\ref{fig:fig3}b, in which the bad cavity regime for the bare system is explored ($\kappa/\omega_m=10$). While standard OM cooling (red circles) is very inefficient, the enhanced optomechanical coupling with a weakly coupled dopant allows for better cooling at red cavity detunings $\Delta_c\sim\kappa/\sqrt{3}$, albeit not to the ground state (blue dots). With a strongly coupled dopant with $g/\omega_m=4$ -- such that effective sideband resolution is achieved for the dressed system ($\Delta_0\sim -1.5\kappa$) -- ground state cooling is possible for blue detunings around $\Delta_c\sim\Delta_0$ (yellow squares).

\subsection{Polariton optomechanics}

\begin{figure}
\centering
\includegraphics[width=0.5\textwidth]{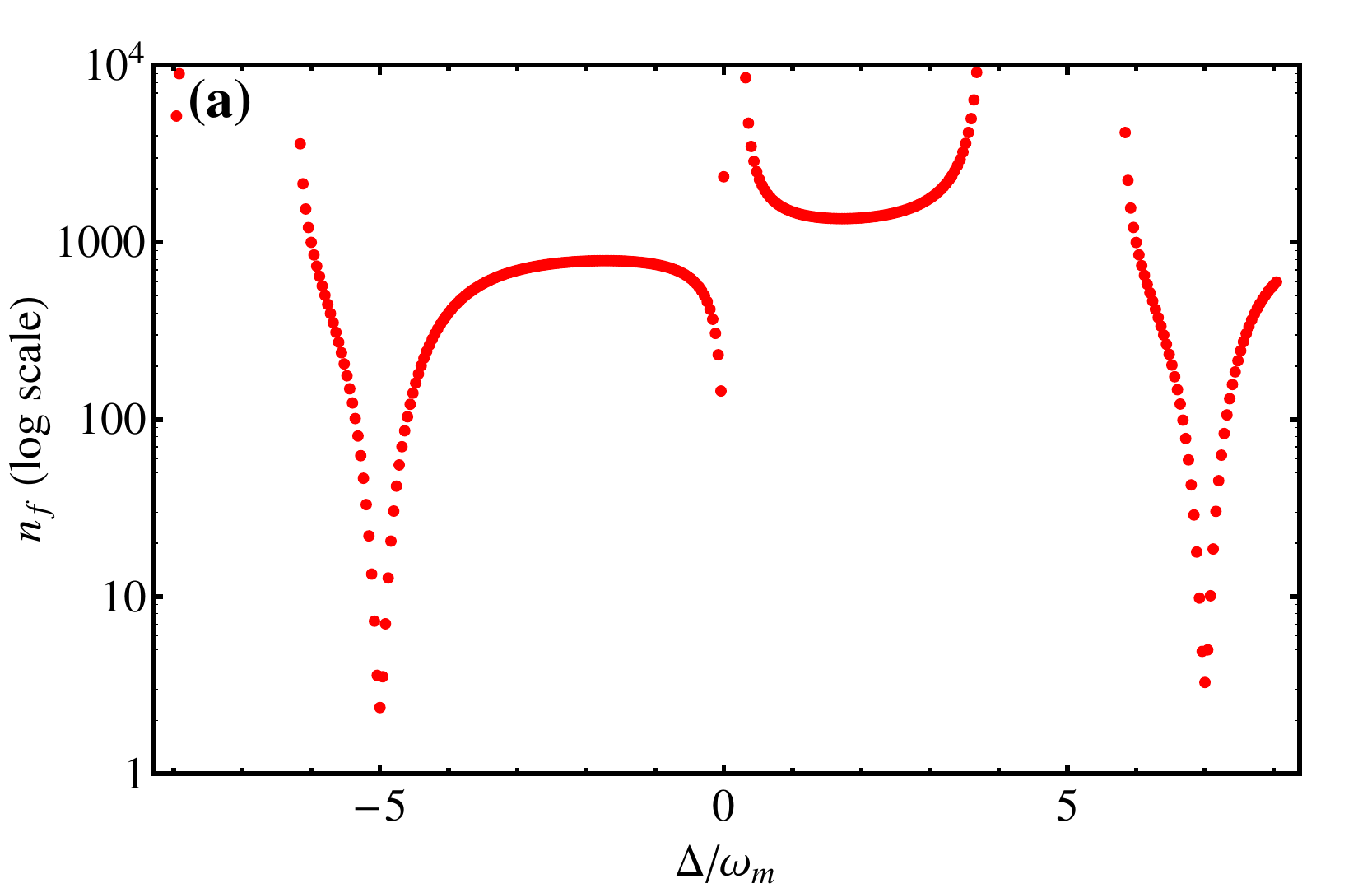}\\
\includegraphics[width=0.5\textwidth]{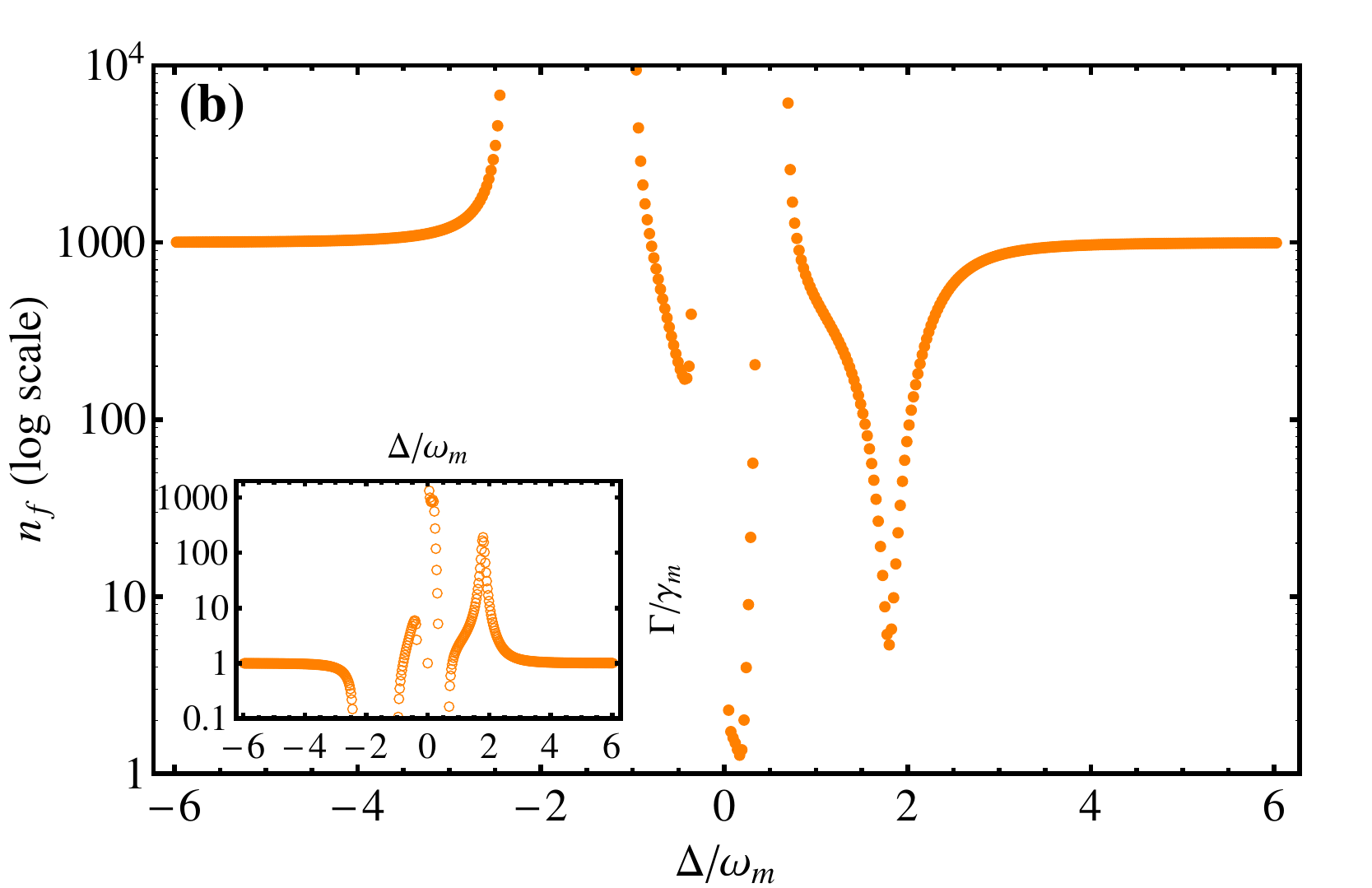}\\
\includegraphics[width=0.5\textwidth]{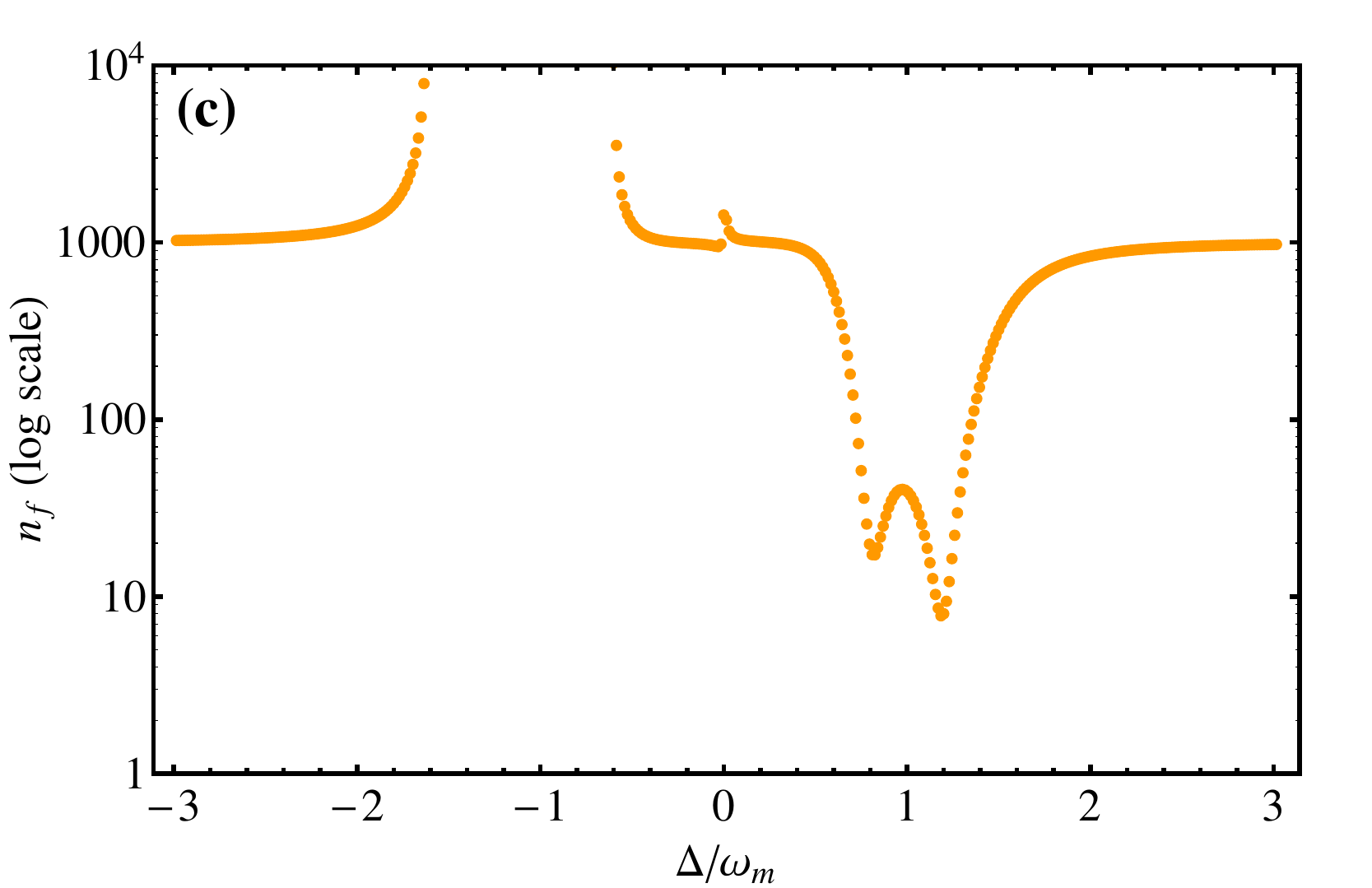}\\
\caption{Final occupation number $n_f$ as a function of polariton detuning $\Delta$, for various coupling strengths: (a) $g/\omega_m=6$, (b) $g/\omega_m=0.8$ and (c) $g/\omega_m=0.2$. Parameters: $\kappa/\omega_m=0.1$, $\gamma/\omega_m=10^{-2}$, $\gamma_m/\omega_m=10^{-5}$, $n_m=10^3$, $G/\omega_m=10^{-2}$. The inset in (b) shows the variation of the optomechanical damping rate $\Gamma$, normalized to $\gamma_m$.}
\label{fig:fig4}
\end{figure}

A natural picture for interpreting these results can be provided by introducing the atom-field mixtures -- \textit{polaritons} -- which diagonalize the Hamiltonian $H_0+H_{int}$. If one assumes for simplicity that the atomic and cavity detunings are kept equal ($\Delta_a=\Delta_c=\Delta$), the polaritons are symmetric combinations of the atomic and field modes
\begin{equation}
\hat{u}=\frac{\hat{a}+\hat{c}}{\sqrt{2}},\hspace{0.2cm}\hat{v}=\frac{\hat{a}-\hat{c}}{\sqrt{2}},
\end{equation}
which give the Hamiltonian
\begin{equation}
H=\frac{1}{2}\omega_m(\hat{p}^2+\hat{q}^2)+[\Delta+g(q_0,\hat{q})]\hat{u}^{\dagger}\hat{u}+[\Delta-g(q_0,\hat{q})]\hat{v}^{\dagger}\hat{v}
\end{equation}
In the strong coupling regime, when the polaritons are well-resolved ($g\gg\kappa,\gamma$), optomechanical cooling is then obtained when one drives the polaritons' red sideband, i.e. when $\Delta=\Delta_{\pm}\equiv\pm g+\omega_m$. A detailed analysis shows that the effective cooling rate is given by $\Gamma=G^2/\bar{\kappa}$, where $\bar{\kappa}=(\kappa+\gamma)/2$ is the effective polariton decay rate. The case $g/\omega_m=6$ is illustrated in Fig.~\ref{fig:fig4}a, which clearly shows optomechanically induced cooling and heating around $\pm g+\omega_m$ and $\pm g-\omega_m$, respectively. A similar behavior is observed in Fig.~\ref{fig:fig4}c for the weak coupling regime $g/\omega_m=0.2$. However, remarkably, heating and cooling are also observed around $\Delta=0$. This may seem surprising as both $\Xi(\omega_m)$ and $\Theta(\omega_m)$ vanish exactly on resonance $\Delta=0$. Nevertheless, for small detunings, the asymmetrical coupling to both polaritons yields a non-zero effective mechanical damping (or antidamping). Indeed, assuming still $\gamma\ll\omega_m$ and expanding $\Xi(\omega_m)$ in Eq.~(\ref{eq:Xi}) at first order in $\Delta$ gives an optomechanical damping/antidamping rate
\begin{equation}
\Gamma\simeq \left(\frac{Gg}{\gamma}\right)^2\frac{4\kappa\Delta_0\Delta}{(\kappa^2-\Delta_0^2+\Delta^2)^2+4\kappa^2\Delta_0^2}
\end{equation}
In the strong ($g\gg\omega_m$) and weak ($g\ll\omega_m$) coupling regimes, $|\Delta_0|\gg 1$ and the damping/antidamping is relatively small $\Gamma\varpropto (Gg/\gamma)^24\kappa\Delta/\Delta_0^3$, as shown in Figs.~\ref{fig:fig4}a and c. However, in the intermediate coupling regime ($g\sim\omega_m$), $\Delta_0$ becomes small and, for $\Delta\simeq\Delta_0$, one retrieves the enhanced cooling rate of Eq.~(\ref{eq:Gammaenhanced}). This is illustrated in Fig.~\ref{fig:fig4}b, where $g/\omega_m=0.8$ and ground state cooling is nearly achieved. The physical interpretation of this somewhat intriguing result is that, when $g\sim\omega_m$ and $\Delta\sim 0$, the lower polariton's red sideband is close to resonance with the upper polariton's blue sideband. The interference between the scattering amplitudes into the two sidebands gives rise to a strong dispersive optomechanical interaction, causing the observed enhanced optomechanical cooling/heating. Note that this effect does not originate from interference in the effective noise terms, but in the scattering amplitudes in the mechanical sidebands (imaginary part of the effective mechanical susceptibility). This is corroborated by the inset Fig.~\ref{fig:fig4}b, showing a variation of $\Gamma$ with the detuning which perfectly correlates with the variation of the final occupation number in the mechanics.

\section{Conclusion}
We have investigated a hybrid approach to optomechanics in which the
addressing of a mechanical resonator's motion is achieved indirectly
via the coupling of light with an embedded dopant comprised of an
ensemble of TLS. We have shown that enhanced effective
optomechanical interactions in the linearized regime can be achieved
with both weakly- or strongly-coupled dopant. As an example we have
shown that such interactions can be used to facilitate ground state
cooling of mechanical modes of resonators for which direct coupling
with light via radiation pressure is otherwise weak.

From a fundamental point of view, it is worth noting that the coupling studied in the present work, in contrast with similar schemes~
\cite{Restrepo2014}, is not provided by an effective two-body interaction
where atom-mechanics coupling is obtained after tracing over the
mediating field, but rather by an intrinsic tripartite interaction where
mechanical operators are directly coupled to dressed light-matter
states [Eq.~\ref{eq:Hint}]. For strong coupling between dopant and light, for instance, interesting dynamics between the mechanics and light-matter polaritons can be engineered.

From a more practical point of view, this approach can be beneficial for the design of mechanical
resonators with e.g. low reflectivity -- such as membranes~\cite{Thompson2008}, levitated submicron particles~\cite{Giesler2012,Asenbaum2013,Kiesel2013,Millen2013}, molecule-embedded polymer layers~\cite{Schwartz2011,Shalabney2014}, etc. The optimization of their mechanical properties can then to some extent be disentangled from the optical ones, since, by choosing the dopant, optical properties can be independently tailored. Moreover, the flexibility in matching doping patterns with vibrational and optical mode profiles
may present a great advantage for multimode addressing of the system.

\section*{Acknowlegments}
We acknowledge support from the EU (ITN CCQED), the Danish National
Council for Independent Research (Sapere Aude program), the Institut
Fran\c{c}ais du Danemark (IFD2013 program), the ERC-St Grant ColdSIM
(No. 307688), EOARD, the Universit\'{e} de Strasbourg through Labex
NIE and IdEX, the JQI, the NSF PFC at the JQI, Initial Training
Network COHERENCE and from the Austrian Science Fund (FWF) via
project P24968-N27.


\begin{thebibliography}{99}

\bibitem{Aspelmeyer2013} M. Aspelmeyer, T. J. Kippenberg, and F. Marquardt, arxiv:1303.0733 (2013).

\bibitem{Treutlein2012} P. Treutlein, C. Genes, K. Hammerer, M. Poggio, and P. Rabl, arxiv:1210.4151 (2012).

\bibitem{Wallquist2009} M. Wallquist, K. Hammerer, P. Rabl, M. Lukin and P. Zoller, Phys. Scr. {\bf T137}, 014001 (2009).

\bibitem{Tian2004} L. Tian and P. Zoller, Phys. Rev. Lett. \textbf{93}, 266403 (2004).

\bibitem{Favero2008} I. Favero and K. Karrai, New J. Phys. {\bf 10}, 095006 (2008).

\bibitem{Singh2008} S. Singh, M. Bhattacharya, O. Dutta, and P. Meystre, Phys. Rev. Lett. \textbf{101}, 263603 (2008).

\bibitem{Hammerer2009b} K. Hammerer, M. Wallquist, C. Genes, M. Ludwig, F. Marquardt, P. Treutlein, P. Zoller, J. Ye, and H. J. Kimble, Phys. Rev. Lett. \textbf{103}, 063005 (2009).

\bibitem{Puller2013} V. Puller, B. Lounis, and F. Pistolesi, Phys. Rev. Lett. {\bf 110}, 125501 (2013).

\bibitem{Restrepo2014} J. Restrepo, C. Ciuti, and I. Favero, Phys. Rev. Lett. {\bf 112}, 013601 (2014).

\bibitem{Meiser2006} D. Meiser and P. Meystre, Phys. Rev. A {\bf 73}, 033417 (2006).

\bibitem{Treutlein2007} P. Treutlein, D. Hunger, S. Camerer, T. W. H\"{a}nsch, and J. Reichel, Phys. Rev. Lett. \textbf{99}, 140403 (2007).

\bibitem{Genes2008} C. Genes, D. Vitali and P. Tombesi, Phys. Rev. A \textbf{77}, 050307(R) (2008).

\bibitem{Ian2008} H. Ian, Z. R. Gong, Yu-xi Liu, C. P. Sun, and F. Nori, Phys. Rev. A {\bf 78}, 013824 (2008).

\bibitem{Genes2009} C. Genes, H. Ritsch and D. Vitali, Phys. Rev. A \textbf{80}, 061803 (2009).

\bibitem{Hammerer2009a} K. Hammerer, M. Aspelmeyer, E. S. Polzik, and P. Zoller, Phys. Rev. Lett. \textbf{102}, 020501 (2009).

\bibitem{Bhattacherjee2009} A. B. Bhattacherjee, Phys. Rev. A {\bf 80}, 043607 (2009).

\bibitem{Hammerer2010} K. Hammerer, K. Stannigel, C. Genes, P. Zoller, P. Treutlein, S. Camerer, D. Hunger, and T. W. H\"{a}nsch, Phys. Rev. A. \textbf{82}, 021803 (2010).

\bibitem{Hunger2010} D. Hunger, S. Camerer, T. W. H\"{a}nsch, D. K\"{o}nig, J. P. Kotthaus, J. Reichel, and P. Treutlein, Phys. Rev. Lett. {\bf 104}, 143002 (2010).

\bibitem{Paternostro2010} M. Paternostro, G. De Chiara, and G. M. Palma, Phys. Rev. Lett. {\bf 104}, 243602 (2010).

\bibitem{Camerer2011} S. Camerer, M. Korppi, A. J\"{o}ckel, D. Hunger, T. W. H\"{a}nsch, and P. Treutlein, Phys. Rev. Lett. {\bf 107}, 223001 (2011).

\bibitem{Genes2011} C. Genes, H. Ritsch, M. Drewsen, and A. Dantan, Phys. Rev. A {\bf 84}, 051801 (2011).

\bibitem{Vogell2013} B. Vogell, K. Stannigel, P. Zoller, K. Hammerer, M. T. Rakher, M. Korppi, A. J\"{o}ckel, and P. Treutlein, Phys. Rev. A {\bf 87}, 023816 (2013).


\bibitem{WilsonRae2004} I. Wilson-Rae, P. Zoller, and A. Imamoglu, Phys. Rev. Lett. {\bf 92}, 075507 (2004).

\bibitem{Lambert2008} N. Lambert, I. Mahboob, M. Pioro-Ladri\`{e}re, Y. Tokura, S. Tarucha, and H. Yamaguchi, Phys. Rev. Lett. \textbf{100}, 136802 (2008).

\bibitem{Yeo2014} I. Yeo, P.-L. de Assis, A. Gloppe, E. Dupont-Ferrier, P. Verlot, N. S. Malik, E. Dupuy, J. Claudon, J-M. Gerard, A. Auff\`{e}ves, G. Nogues, S. Seidelin, J-Ph. Poizat, O. Arcizet, and M. Richard, Nat. Nanotech. {\bf 9}, 106 (2014).


\bibitem{Rabl2009} P. Rabl, P. Cappellaro, M. V. Gurudev Dutt, L. Jiang, J. R. Maze, and M. D. Lukin, Phys. Rev. B \textbf{79}, 041302(R) (2009).

\bibitem{Arcizet2012} O. Arcizet, V. Jacques, A. Siria, P. Poncharal, P. Vincent and S. Seidelin, Nat. Phys. {\bf 7}, 879 (2012).

\bibitem{Kolkowitz2012} S. Kolkowitz, A. C. Bleszynski Jayich, Q. P. Unterreithmeier, S. D. Bennett, P. Rabl, J. G. E. Harris, and M. D. Lukin, Science {\bf 335}, 1603 (2012).

\bibitem{Zhang2013} Z. Yin, T. Li, X. Zhang, and L. M. Duan, Phys. Rev. A {\bf 88}, 033614 (2013).

\bibitem{Scala2013} M. Scala, M. S. Kim, G. W. Morley, P. F. Barker, and S. Bose, Phys. Rev. Lett. {\bf 111}, 180403 (2013).

\bibitem{Ramos2013} T. Ramos, V. Sudhit, K. Stannigel, P. Zoller, and T. J. Kippenberg, Phys. Rev. Lett. {\bf 110}, 193602 (2013).

\bibitem{OConnell2010} A. D. O'Connell, M. Hofheinz, M. Ansmann, R. C. Bialczak, M. Lenander, E. Lucero, M. Neeley, D. Sank, H. Wang, M. Weides, J. Wenner, J. M. Martinis, and A. N. Cleland, Nature {\bf 464}, 697 (2010).

\bibitem{Pirkkalainen2013} J.-M. Pirkkalainen, S. U. Cho, Jian Li, G. S. Paraoanu, P. J. Hakonen, and M. A. Sillanp\"{a}\"{a}, Nature {\bf 494}, 211 (2013).

\bibitem{Palomaki2013} T. A. Palomaki, J. W. Harlow, J. D. Teufel, R. W. Simmonds, and K. W. Lehnert, Nature {\bf 495}, 210 (2013).


\bibitem{Thompson2008} J. D. Thompson, B. M. Zwikl, A. M. Jayisch, F. Marquardt, S. M. Girvin, and J. G. E. Harris, Nature \textbf{452}, 72 (2008)

\bibitem{Genes2008b} C. Genes, D. Vitali, P. Tombesi, S. Gigan, and M. Aspelmeyer, Phys. Rev. A \textbf{77}, 033804 (2008).

\bibitem{WilsonRae2007} I. Wilson-Rae, N. Nooshi, W. Zwerger, and T. J. Kippenberg, Phys. Rev. Lett. \textbf{99}, 093901 (2007).

\bibitem{Marquardt2007} F. Marquardt, J. P. Chen, A. A. Clerk, and S. M. Girvin, Phys. Rev. Lett. \textbf{99}, 093902 (2007).

\bibitem{Dantan2008} A. Dantan, C. Genes, D. Vitali, and M. Pinard, Phys. Rev. A \textbf{77}, 011804(R) (2008).

\bibitem{Giesler2012} J. Giesler, B. Deutsch, R. Quidant, and L. Novotny, Phys. Rev. Lett. {\bf 109}, 103603 (2012).

\bibitem{Asenbaum2013} P. Asenbaum, S. Kuhn, S. Nimmrichter, U. Sezer, and M. Arndt, Nat. Comm. {\bf 4}, 2743 (2013).

\bibitem{Kiesel2013} N. Kiesel, F. Blaser, U. Delic, D. Grass, R. Kaltenbaek, and M. Aspelmeyer, PNAS USA, {\bf 110}, 14180 (2013).

\bibitem{Millen2013} J. Millen, T. Deesuwan, P. Barker, and J. Anders, arxiv:1309.3990 (2013).


\bibitem{Schwartz2011} T. Schwartz, J. A. Hutchinson, C. Genet, and T. W. Ebbesen, Phys. Rev. Lett. {\bf 106}, 196405 (2011).

\bibitem{Shalabney2014} A. Shalabney, J. George, J. Hutchison, G. Pupillo, C. Genet, and T. W. Ebbesen, arxiv:1403.1050 (2014).





\end{thebibliography}
\end{document}